\documentclass[11pt,twoside]{article}

\usepackage{asp2006}
\usepackage{epsf}
\usepackage{psfig}
\usepackage{lscape}

\begin{document}
\def\etal   {{\it et al. }}                     
\def\cappage #1 #2 #3 {\vfill\eject\pageno=#1
\vglue 10 true in minus 10 true in \noindent{\bf Figure #2.} #3}
\def\ee #1 {\times 10^{#1}}
\def\ut #1 #2 { \, \hbox{#1}^{#2}}
\def\u #1 { \, \hbox{#1}}
\def\lsol{\, \hbox{$\hbox{L}_\odot$}}
\def\msol{\, \hbox{$\hbox{M}_\odot$}}
\def\kms {\, \hbox{km}\,\hbox{s}^{-1}}
\def\persec {\, \hbox{s}^{-1}}
\def\percc {\, \hbox{cm}^{-3}}
\def\persqcm {\, \hbox{cm}^{-2}}
\def\mic {\, \mu \hbox{m}}
\def\half{{\textstyle {1\over 2}}}
\def\thalf{{\textstyle{ 3\over 2}}}

\title{Multi-Wavelength Study of Sgr A*: The Short Time Scale Variability}
\author{F. Yusef-Zadeh$^1$, J.  Miller-Jones$^2$, D. Roberts$^3$, M. Wardle$^4$, M. Reid$^5$, 
K. Dodds-Eden$^6$, D. Porquet$^7$ \& N. Grosso$^7$} 
\affil{$^1$Dept Physics and Astronomy, Northwestern University, Evanston, IL. 60208}  
\affil{$^2$NRAO, Charlottesville, 520 Edgemont Road, VA 22903}
\affil{$^3$Adler Planetarium \& Astronomy Museum 1300 S. LSD,  Chicago, IL 60605}
\affil{$^4$Department of Physics, Macquarie University, Sydney NSW 2109,Australia} 
\affil{$^5$Harvard-Smithsonian CfA, 60 Garden Street,
Cambridge, MA 02138} 
\affil{$^6$Max-Plank-Institut f\"ur Extraterrestrische Physik 1312, D-85471, Garching, Germany} 
\affil{$^7$Observatoire astronomique de Strasbourg, Universit\'e
de Strasbourg, NRS, INSU, 11 rue de l'Universit\'e, 67000 Strasbourg,
France} 

\begin{abstract} 
To understand the correlation and the radiation mechanism of flare
emission in different wavelength bands, we have coordinated a
number of telescopes to observe Sgr~A* simultaneously. We focus only
one aspect of the preliminary results of our multi-wavelength observing 
campaigns,
namely, the short time scale variability of emission from Sgr~A* in
near-IR, X-ray and radio wavelengths.  The structure function analysis
indicate most of the power spectral density  is detected on 
hourly time 
scales in all
wavelength bands. We also report minute time scale variability at 7
and 13mm placing a strong constraint on the nature of the variable
emission.  The hourly time scale variability can be explained in the 
context
of a model in which the peak frequency of emission shifts toward
lower frequencies as a self-absorbed synchrotron source expands
adiabatically near  the acceleration site.  The short time scale
variability, on the other hand, places a strong constraint on the size
of the emitting region.  Assuming that rapid minute time scale
fluctuations of the emission is optically thick in radio wavelength,
light travel arguments requires relativistic particle energy, thus
suggesting the presence of outflow from Sgr~A*.
\end{abstract}


\section{Introduction}   

Observations of stellar orbits in the proximity of the enigmatic radio
source Sgr~A* located at the very dynamical center of our galaxy 
(Reid and Brunthaler 2004) have
established that it is a 4 $\times 10^6$\msol\ black hole (Ghez et al.
2008; Gillessen et al. 2009). 
The relative proximity of Sgr~A* compared to AGNs allows us to observe the 
spectacular
activities of the central engine at remarkably small spatial scales.
The variable continuum flux of AGNs is known to signal activities of
the central engine and is detected throughout the entire
electromagnetic spectrum with time scales  ranging from days to 
years
(e.g., Arshakian et al. 2010).  The bulk of the continuum flux is
believed to be generated in the accretion disk, where the localization
of the source of variable continuum emission becomes essential for our
understanding of the launching and transport of energy in active
galaxies.  Sgr~A* provides the best laboratory to study the properties
of low-luminosity accreting black holes.  The emission from Sgr~A* is
therefore the subject of intense scrutiny, which will have a
long-lasting impact on our understanding of the radiatively
inefficient accretion flow into, as well as outflows from, massive
black holes.  The time scale for variability is proportional
to the mass of the black hole, thus  Sgr~A*, which is a hundred
times closer to us than the next nearest example, presents an
unparalleled opportunity to closely study the processes by which gas is
captured, accreted or ejected, by characterizing the time variability
over scales of minutes to years.

The luminosity of Sgr~A* is thought to be due to partial capture of thermal winds from a neighboring      
cluster of massive stars.  However, the bolometric luminosity of Sgr~A* ($\sim 100$ \lsol) is several
orders of magnitude lower than expected for the estimated accretion rate, prompting a number of
theoretical models to explain its very low radiative efficiency and matching the spectral energy
distribution (SED) of its quiescent emission. 
 The broad-band spectrum of Sgr~A* peaks at sub-millimeter wavelengths, which is the dividing line
between optically thin infrared and optically thick millimeter and radio emission.  The existence of flare
emission from Sgr~A* is now well established in both optically thick and thin regimes.  What is really
exciting is that we are beginning to peer into the closest supermassive black hole 
and are discovering rapid, time 
variability that likely stems from gas dynamical flow and radiation 
emitted very near the event horizon.

\section{The Variability of Sgr A*}   
\subsection{Theoretical Simulations}

The emission from Sgr~A* consists of quasi-steady and variable
components (e.g., Hornstein et al. 2007).  A variety of models have
been proposed to explain the steady emission from Sgr~A* by fitting
its SED -- for example a thin accretion disk, a disk and jet, outflow,
an advection-dominated accretion flow, radiatively inefficient
accretion flow, and advection-dominated inflow/outflow solutions.
These studies have not been able to uniquely identify the underlying
mechanism to explain the underluminous nature of Sgr~A*.  However,
more recently, detailed MHD simulations of the inner disk of
radiatively inefficient flow of a few Schwarzschild radius R$_s$, have
found  low-level flux variations (``quiescent'' emission) and
strong transient (``flare'') emission in almost all wavelength bands
(e.g., Goldston et al. 2005; Chan et al.  2009; Moscibrodzka et al.
2009).  These simulations have become increasingly sophisticated by
adding more physics, such as GR effects and a Kerr rather than a
Schwarzschild black hole.  In particular, these theoretical studies
investigate  the structure and the physical parameters of the hot
synchrotron emitting plasma in the vicinity of Sgr~A* at near-IR (NIR)
wavelengths and predict  long-term evolution of its emission.
For example, the variability of emission on a 100 and 1000-hour time
scale was predicted from radiatively inefficient accretion flow.
Goldston et al. (2005) argue that the main reason for the time
variability of the quasi-steady component is due to changes in the
magnetic field.

In a more recent MHD simulation study, Chan et al. (2009) find that
MHD turbulence can only produce quiescent variability of a factor of
two in the luminosity of Sgr~A*.  The strong NIR flaring of Sgr~A* is
then explained by the external material from the outer disk raining
down on the inner disk, thus producing transient outbursts, as seen in
NIR and X-ray wavelengths.  In this picture, strong but short term
QPOs are predicted to emerge in the simulated light curves. Lastly, as
we heard from C. Gammie and F. Yuan in this conference, Moscibrodzka
et al.  (2009) have constructed radiative models of Sgr~A* and have
included the black hole spin.  Their study is most consistent with a
black hole spin of a=0.9, where a is the spin parameter, as it is
likely to show time-variable emission in almost all wavelength bands.

\begin{figure}
\center
\center
\plotfiddle{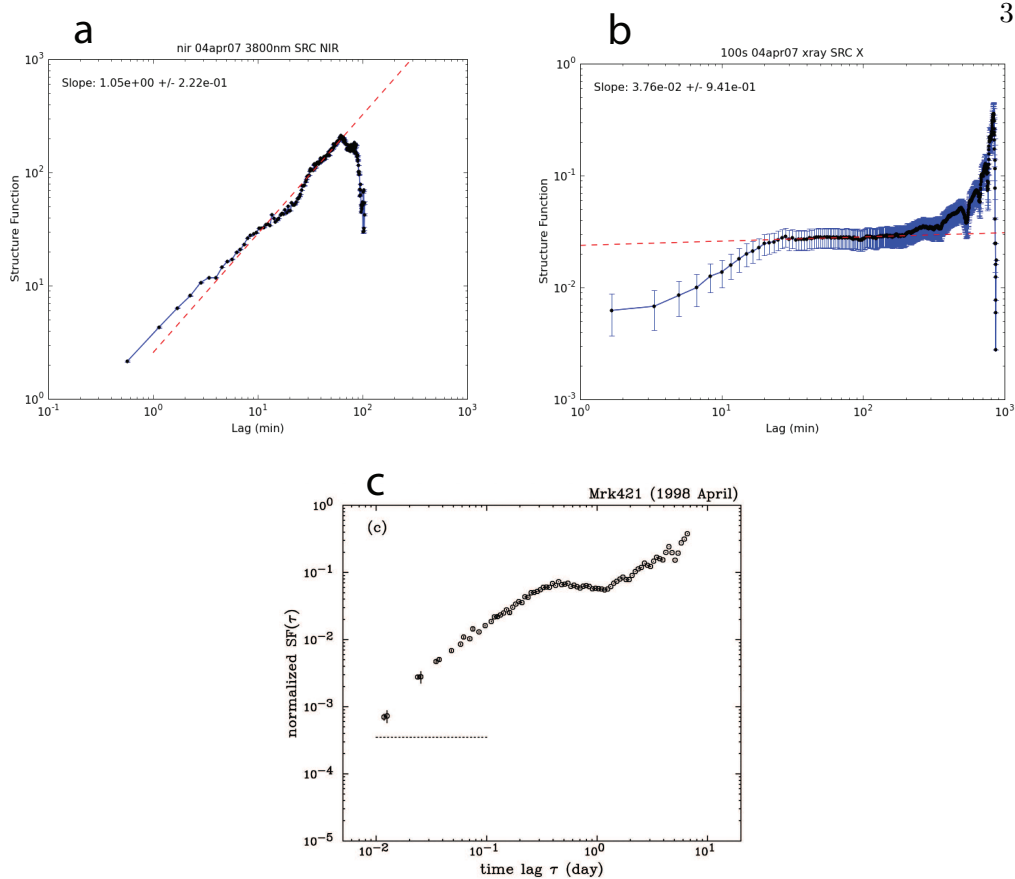}{4.075in}{0}{68.0}{68.0}{-200}{0}
\caption{
(a - Top Left) 
A structure function plot for a day of observation
on 2007, April 4 using VLT at 3.8$\mu$m (Dodds-Eden et al. 2009). 
The power law fit is shown by a dashed line. 
(b - Top Right) 
Similar to (a) except the data were taken with XMM
on the same day but longer observation. 
There was considerable activity on this day in both NIR and X-ray wavelengths
(Porquet et al. 2008).
(c - Bottom)  An X-ray  structure function plot of Mrk 421 is reproduced 
from Fig 4c of Kataoka et al. (2001). }
\end{figure}

There are indeed many MHD simulations of accretion disks around
massive black holes, all having their own limitations and
assumptions. Given the limitations and uncertainties in each simulated
light curve, these studies can only be demonstrative if the
statistical properties of the simulated light curves are in reasonable
agreement with the statistics drawn from the observed light curve of
Sgr~A*.  Variability of the NIR emission of Sgr~A* is a key carrier of
information regarding the dynamics of the accretion flow.  Short
($\sim$minutes) time scales probe the transient acceleration of
particles associated with the dissipation of kinetic and magnetic
energy on small scales within the accretion flow.  Intermediate time
scales (hours) probe the dynamical evolution of the accretion flow on
a timescale of a few orbits, as in the theoretical study of Chan et
al.\ (2009).  Longer time scales (days to months) probe the crucial
and unexplored dynamics of the stochastic feeding of the inner
accretion flow from larger radii and the draining of the inner flow
onto the black hole. Future variability measurements of Sgr~A* are
expected to show the low-level accretion flow that produces quiescent
flux variations of a factor of 2 whereas the flaring activity from
transient events is expected to show flux variations of about a factor ten.

\begin{figure}
\plotfiddle{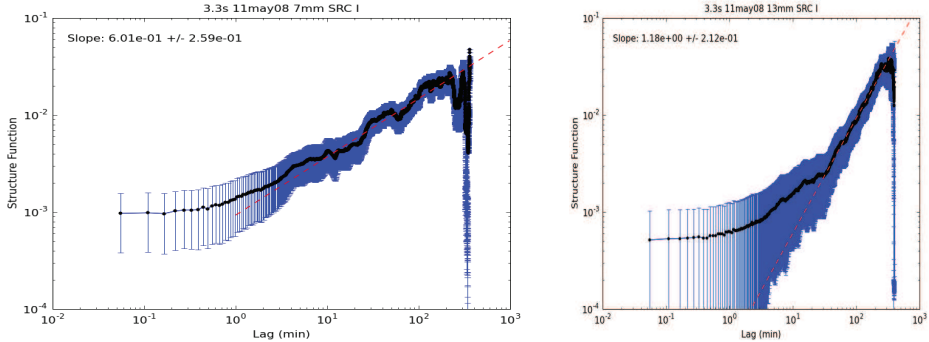}{1.8in}{0}{60.0}{50.0}{-180}{0}
\caption{(a - Left) A structure function plot for VLA observation of
Sgr~A* at 7mm on 2008, May 11 with data sampling of 3.3sec. 
 (b -
Right) Similar to (a) except at 13mm. Both 7 and 13mm data are taken
simultaneously. The error bars are shown in blue whereas 
 the power law fit is shown in dashed red lines.  
The amplitude are in  Jy$^2$. 
The squares of the mean measurement 
errors are 6 and 8$\times10^{-4}$ Jy$^2$ at 7 and 13mm, respectively.} 
\end{figure}

\subsection{Structure Function Analysis of NIR and X-ray  Variability} 

Previous NIR measurements have characterized the intrinsic variability
time scale of Sgr~A* by using structure function analysis, which is
widely used in the study of AGN light curves in X-rays (e.g., Kataoka
et al. 2001). The structure function (SF) is the mean difference of
pairs of flux measurement separated by time lag $\tau$, i.e.  $<[S(t)
- S(t+\tau)]^2>$ (Simonetti et al. 1985).
 The slope of the structure function $\beta$ (where
SF$\propto\tau^{\beta}$) is determined by the slope of the power
spectral density. It turns out that the slope of the structure
function is related to the index of the power spectrum of fluctuations
$\alpha$ where power spectrum P $\propto f^{-{\alpha}}$.  The relationship between
$\alpha$ and $\beta$ is approximately $\alpha\sim\beta+1$ when $\alpha$ is greater  than 
or equal one (Kataoka et al. 2001). 
 Flicker noise
with $\alpha\sim1$ or $\beta$=0 has been identified in the power spectrum
of many AGNs. Breaks in the power spectra are seen in both AGN and
X-ray binaries, with the break timescale scaling with black hole mass
and bolometric luminosity (i.e. mass accretion rate), McHardy et
al. (2006).

Structure function analysis has also been carried out with several
nights of Keck observations of Sgr~A* (Do et al.  2009). 
Different nights of Keck observations (Do et al. 1990) show 
values of $\beta$ varying between 0.26 and 1.37 with lag times ranging between 1 to 40 minutes.  
A follow-up work by Meyer et al. (2009) investigated the long time range of up to four years to 
determine the turnover in the structure function analysis of NIR data. 
These authors find   a slope of 2.1 with a characteristic time scale for the  turnover at 
around 154 min, though with 
a lage  uncertainity due to the poor sampling of NIR data on  long time scales. 

We have used data taken with VLT and XMM observations on 2007, April 4 at NIR and X-ray 
wavelengths.  Figures 1a,b show the structure function plots of NIR and X-ray observations that 
are sensitive only to time lags ranging between  30sec to $\sim$100min and 100sec to 800min 
for the VLT  and XMM data, respectively.  The 
power-law fit to the NIR data implies slope of 1.05$\pm0.22$ in the structure function 
suggesting random walk noise. 
The value of $\beta$ from VLT measurements is  consistent with Keck measurements presented by 
Do et al. (2009) but difffer  with the value of $\beta$=2.1 determined from combining 
VLT and Keck data for four years (Meyer et al. 2009).  
 The turnover in the  VLT structure function is seen at $\sim70$min.  
The turnover  
is either due to time lags that are not sampled well at large time lags or due to 
a lack of  correlation between signals with different time lags. 

We also note variability at less than 1 
minute at NIR wavelengths (see also the analysis by Dodds-Eden et al. 2010 and Do et al. 2009). 
There is no evidence for QPOs in the time domain that was searched which is consistent with 
the results given by Do et al. (2009) and Meyer et al. (2008).  
The nature of QPO activity 
of a hot spot orbiting Sgr~A* is hotly debated mainly because of the low signal-to-noise and 
possibly due to an intermittent nature of such a behavior (e.g., Eckart et al. 2006).

Unlike the structure function plot of NIR data which is fit a by a single power law, 
longer duration of XMM observations, as shown in  in Fig. 1b,  give different characterization of
the variability of Sgr~A* in X-rays.  We note a steep rise 
at short time scales,  similar to that of NIR, followed by 
a plateau with time lags ranging between  30 and 300 minutes 
 before a steepening of the structure function is noted 
again  at longer time scales. 
Due to limited sensitivity and 100-second sampling of  X-ray data, 
the flat part of the SF with time lags of  few minutes indicate 
that there is no  detectable minute time scale 
variability and that the emission is dominated by  white noise.    The plateau 
time lags range 
between 30 and 300 minutes  is also consistent with a white noise where there is 
no correlation of signals.
At time lags greater than 300 minutes, the correlation begins again and most 
of the power appears to be in this time domain.  
The X-ray shape of the structure function plot of Sgr A*  is remarkably similar to 
that  of Mrk 421 except that the time lags of Mrk 421 are  an order of 
magnitude 
larger than that detected for Sgr A*. For comparison, Figure 4c shows 
a structure function for the light curve 
of  Mrk 421 measured by ASCA (Kataoka et al. 2001).


\begin{figure} 
\plotfiddle{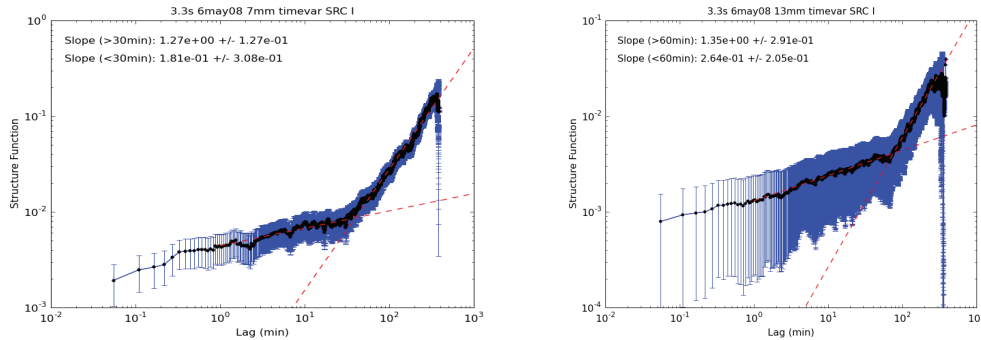}{1.6in}{0}{65.0}{65.0}{-200}{0} 
\caption{ (a - Left) A structure function 
plot for VLA observation of Sgr~A* at 7mm on 2008, May 6 with data sampling of 3.3sec.  (b - 
Right) Similar to (a) except at 13mm. The error bars for individual amplitude are shown in blue 
and the power law fits are indicated in red dashed lines. 
The squares of the mean measurement 
errors are 7 and 9$\times10^{-4}$ Jy$^2$ at 7 and 13mm, respectively.} 
\end{figure}


\subsection{The Structure Function Analysis of Radio Variability}

Previous light curve analysis of radio data at 7mm have shown that the
duration of typical radio flares is about two hours, which is similar
to the duration of flares observed at sub-mm 
wavelengths (Yusef-Zadeh et al. 2006; Marrone et al. 2006). This is consisent with 
the power spectrum analysis of the time variability of Sgr A* which suggested 
intraday variability at 3mm (Mauerhan et al. 2005). 
Based on the duration of radio  flares, it was argued
that cooling could be due to adiabatic expansion, with the implication
that flare activity can be caused by an expanding plasma blob
(Yusef-Zadeh et al. 2006).  The measurements in support of this
picture indicated the peak frequency of the emission (e.g., the
initial optically thin near-IR flare) shifts toward lower frequencies
(sub-millimeter, millimeter and then radio) as a self-absorbed
synchrotron source expands adiabatically near the acceleration
site.  To investigate the short time scale variability, we have
constructed several structure function plots of Sgr~A* from radio
variability data taken with the VLA mainly from a multi-wavelength
observing campaign that took place in April and May 2008.  We present
three sets of plots from three different days of VLA observations.
 
Figures 2a,b show the structure function plots of Sgr A* 
 indicating  time scales ranging roughly between
1  to 300 minutes 
in the variability of Sgr~A* 
at 7 and 13mm on 2008, May 11,
respectively. Most of the power in the time variability 
is detected between one and three hour
time scales, consistent with earlier light curve analysis.  The fit to
the slope of  the structure function at long time scales is  
a   power law  giving  values
0.60$\pm0.26$ and 1.18$\pm$0.21 at 7 and 13mm, respectively.  On the
short time scale, we note a shallower slope of the structure function
with large error bars indicating that there is no minute time scale
variability in radio wavelength on this day. This is consistent with 
measurements errors in amplitude (white noise). 

To examine the frequency dependence of the variability of Sgr~A*,
Fig. 3 shows the structure function plots based on data taken
simultaneously on 2008, May 8 at 7 and 13mm, respectively.  Both plots
show the presence of minute time scale variability going down to
$\sim$0.1 minute.  The power law fit to the structure function give different
values for short and long time scales; the power law fit to the slope of the 7
and 13mm data at short time scales gives values 0.18$\pm0.31$ and
0.26$\pm0.20$ whereas the power law fit to the long time scale slope, are
estimated to be 1.27$\pm$0.13 and 1.35$\pm$0.29, respectively.
Furthermore, the transitions from a shallow to a steep slope are 30
and 60 minutes at 7 and 13mm, respectively.  The long time variability
is identified with the typical two hour duration of flaring events in
radio wavelengths.  The transition time scales from 30 and 60 minutes
at these two different wavelengths are likely to be due to optical
depth effect in an adiabatic expanding synchrotron source.  Previous
time delay measurements which show a time delay of 20-40 minutes
between the peaks of emission at 7 and 13mm (Yusef-Zadeh et al. 2009) 
as well as the stretching of the duration of the flare emission in the  
time  domain 
at lower frequencies. The latter effect is 
likely to be responsible for  the transition in the 
slope seen in the structure 
function analysis.

\begin{figure}
\plotfiddle{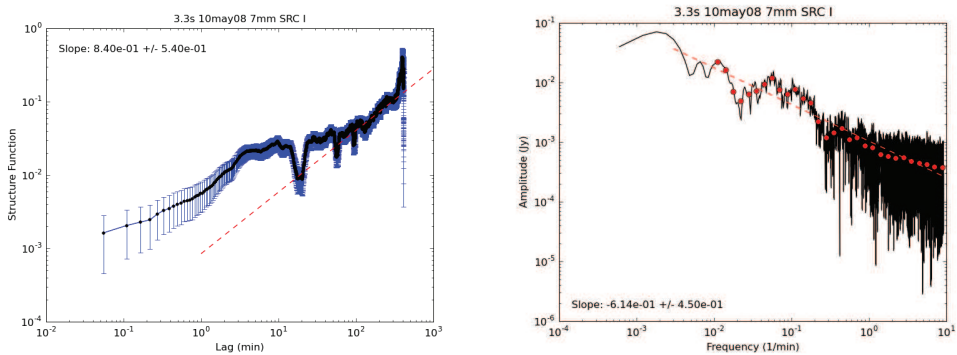}{1.8in}{0}{70.0}{70.0}{-200}{0}
\caption{(a - Left) A structure function plot for VLA observation of
Sgr~A* at 7mm on 2008, May 10 with data sampling of 3.3sec. 
The squares of the mean measurement 
errors are 7$\times10^{-4}$ Jy$^2$ at 7mm.
 (b - Right) The corresponding power spectral density of (a).
The red dots show binned data when averaged for  a given lagtime.
The slope to both plots are presented as dashed lines. 
}
\end{figure}

Lastly, Figure  4 shows another structure function plot and the
corresponding power spectral density  based on data taken
on 2008, May 10.  The shape of the structure function indicates
multiple slopes.  A power law fit to the slope at long time scales gives a value
of 0.84$\pm$0.54 which is consistent with typical slopes of other
structure functions shown in Figures 2 and 3.  What is new in Figure 4a is
the appearance of a number of dips in the structure function
indicating a possible QPO activity.  The strongest dip is seen around
20 minutes followed by weaker dips at 50 and 100 minutes.  Similar
dips are also noted in the structure function at 13mm taken
simultaneously with the 7mm data.  The power spectral density, 
as presented in Fig. 4b, gives another presentation of
a possible evidence of QPO activity. 
 The best-fitting power laws appear to have power law index of 0.6
which falls somewhere between white noise  and
flicker noise. Deviations from a power law fit to the power structural density 
could be the result of dips detected in the structure function plot.


\section{Discussion}

Our multi-wavelength  monitoring of Sgr A* 
characterizes  the intrinsic time
variability of Sgr~A* by studying structure function of the observed light 
curves. 
Unlike near-IR structure function plots that can be fit by 
single power laws, radio and X-ray structure function plots
are fit by multiple power laws in different time lag domains. 
Structure function analysis of the data presented in near-IR, X-ray
and radio data suggest that most of the power falls in the long time
scale fluctuation of the emission from Sgr~A* and no obvious QPO activity. 
We also note
short time scale variability at these wavelengths when the data are 
sampled properly. In particular, the minute-time scale variability 
is best seen at NIR wavelengths in Fig. 1a.
The  structure function also  goes down well below one minute 
at radio wavelengths showing a slope which is 
not completely flat at the shortest lags. 
The best case for such minute time scale variability is seen in Figure 3a 
at 7mm  where the slope is not consistent with being flat for lag times
less than 0.3 minute.  The square of the mean of the measurement errors for 
3.3sec sample data at 7mm 
is $\sim0.0007$ Jy$^2$ which is more than a factor of two  less than the   amplitude of the 
structure 
function 2$\times10^{-3}$ Jy$^2$ at the shortest lag time. 


Assuming that the sub-minute time scale variability in NIR and radio 
wavelengths are  significant, 
they place a strong constraint on 
the size of
the region from which variable emission arises.  
Using light travel time arguments, the
minute-time scale variability gives an upper limit to the scale length
of radio variation $\sim$1/8 AU. 
For comparison, the Schwarzschild
radius of Sgr~A* is 10 $\mu$as or 0.1 AU at the distance of 8 kpc
(Doeleman et al. 2008).  
Assuming that the emission is
optically thick at 7mm, then the inferred brightness temperature is
$> 5\times10^{10}$K corresponding to particle energy $> 4.5$MeV.  These
high energy particles have relativistic speeds and can in principle
leave the gravitational potential of Sgr~A* in the form of outflow.
The inferred scale length corresponding to one-minute light travel time  
is comparable to  to the time averaged spatially
resolved 0.1AU  scale  observed at 1.3mm by Doeleman et al. (2008).
The quiescent variable emission from Sgr~A* could then be interpreted
mainly as an ensemble average of numerous flares that are detected on
minute-time scale.  Alternatively, the short time scale emission or
quiescent variability could be due to fluctuations in the accretion
flow of Sgr~A* due to magnetic field fluctuations resulting from MRI,
as recent MHD simulations in a number of studies indicate.  A more
detailed account of these results will be given elsewhere.





\section{References}

Arshakian, T.~G., Leon-Tavares, J., Lobanov, A. P.
et al. 2010, MNRAS, 401, 1231\\
Chan, C.-K., Liu, S., Fryer, C.~L., Psaltis, D., Ozel, F.,
Rockefeller, G. and Melia, F. 2009, ApJ, 701, 521\\
Doeleman, S.~S., Weintroub, J., Rogers, A.~E.~E., Plambeck, R.,
Freund, R.,  Tilanus, R.~P.~J. et al.  2008, Nature, 455, 70\\
Ghez, A.~M., Salim, S., Weinberg, N.~N., Lu, J.~R., Do, T.,
Dunn, J.~K., Matthews, K., Morris, M.~R., Yelda, S.,
Becklin, E.~E. and Kremenek, T. et al. 2008, ApJ, 689, 1044\\
Hornstein, S.~D., Matthews, K., Ghez, A.~M., Lu, J.~R.,
et al. 2007, ApJ, 667, 900\\
Eckart, A., Schodel, R.,  Meyer, L.,  Trippe, S. Ott, T.
and Genzel, R. 2006, A\&A, 455, 1\\
Gillessen, S., Eisenhauer, F. et al. 2009, ApJ, 692, 1075\\
Goldston, J.~E., Quataert, E. and Igumenshchev, I.~V. 2005,
ApJ, 621, 785\\
Kataoka, J., Takahashi, T., Wagner, S.~J. et al.
2001, ApJ, 560, 659\\
McHardy, I.~M., Koerding, E., Knigge, C.,  Uttley, P. et al.
2006, ApJ, 444, 730\\
Mauerhan, J. C., Morris, M., Walter, F. \& Baganoff, F. K. 2005, 623, L25\\ 
Meyer, L., Do, T., Ghez, A., Morris, M. R., et al. 2009, ApJ, 694, L87\\
Meyer, L., Do, T., Ghez, A., Morris, M. R., et al. 2008, ApJ, 688, L17\\
Moscibrodzka, M., Gammie, C.~F., Dolence, J.~C.
2009, ApJ, 706, 497\\
Porquet, D.,  Grosso, N., Predehl, P.,
Hasinger, G., Yusef-Zadeh, F.,  Aschenbach, B.,
et al.  2008, A\&A, 488, 549\\
Reid, M. J. and Brunthaler, A. 2004, ApJ, 616, 872\\ 
Simonetti, J. H., Cordes, J. M. and Heeschen, D. S.
1985, ApJ, 296, 46\\
Yusef-Zadeh, F., Roberts, D.,  Wardle, M., Heinke, C.~O.
and Bower, G.~C.  2006, ApJ, 650, 189\\
Yusef-Zadeh, F.,  Bushouse, H.,  Wardle, M.,  Heinke, C.
et al.  2009, ApJ, 706, 348\\

\end{document}